\documentclass{PoS}
\usepackage[numbers,sort&compress]{natbib}
\usepackage{natbibspacing}

\usepackage{eucal}

\newcommand\para{\parallel}

\title{The form factors for $B \to \pi l \nu$ semileptonic decay from $2+1$ flavors of domain-wall fermions}

\ShortTitle{The $B\to \pi l \nu$ form factors with domain-wall fermions}

\author{\speaker{Taichi Kawanai}\\
  Theoretical Research Division, Nishina Center, RIKEN, Wako 351-0198, Japan\\
  Physics Department, Brookhaven National Laboratory, Upton, NY 11973, USA\\
  RIKEN-BNL Research Center, Brookhaven National Laboratory, Upton, NY 11973, USA\\
  E-mail: \email{taichi.kawanai@riken.jp}}

\author{Ruth S.~Van de Water\\
	Theoretical Physics Department, Fermi National Accelerator Laboratory, Batavia, IL  60510, USA\\
        E-mail: \email{ruthv@fnal.gov}}

\author{Oliver Witzel\\
        Center for Computational Science, Boston University, 3 Cummington Mall, Boston, MA 02215, USA\\
	E-mail: \email{owitzel@bu.edu}
      }

\abstract{
  We present a calculation of  the $B\to \pi l\nu$ form factors with domain-wall light quarks and 
  relativistic $b$-quarks on the lattice.
  We work with the $2 + 1$ flavor domain-wall fermion and 
  Iwasaki gauge-field ensembles generated by the RBC and UKQCD Collaborations.
  The chiral-continuum extrapolation is performed using SU(2) hard-pion chiral perturbation theory.
  To extrapolate the lattice form factors to the full kinematic range, 
  we use the model independent $z$-parameterization and impose the kinematic constraint $f_+(0) = f_0(0)$ 
  at zero momentum transfer.
}

\FullConference{31st International Symposium on Lattice Field Theory - LATTICE 2013\\
		July 29 - August 3, 2013\\
		Mainz, Germany}

\begin{document}

\section{Introduction}
The precise determination of the Cabibbo-Kobayashi-Maskawa~(CKM) 
matrix element $|V_{ub}|$ provides a strong test of the standard model.
The lattice calculation of the hadronic form factor $f_+$
plays an essential role in the determination of $|V_{ub}|$ 
from  $B \to \pi l\nu$ exclusive semileptonic decay.
When combined with the hadronic form factor $f_+(q^2)$, 
the value of $|V_{ub}|$ can be obtained by experimental measurements of the differential decay rate via 
\begin{equation}
 \frac{d\Gamma(B\to \pi l\nu)}{dq^2} = 
\frac{G_F^2|V_{ub}|^2}{192\pi^3 m_B^3}\left[
  (m_B^2+m_\pi^2-q^2)^2-4m_B^2m_\pi^2
\right]^{3/2} |f_+(q^2)|^2,
\end{equation}
where the momentum transfer $q^\mu\equiv p_B^\mu - p_\pi^\mu$ and
we neglect the mass of the outgoing lepton.
The form factors $f_+$ and $f_0$
parameterize the hadronic element of the $b\to u$ vector current
 $\mathcal{V}^\mu\equiv i\bar{u}\gamma^\mu b$:
\begin{equation}
 \langle  \pi | \mathcal{V}^\mu | B \rangle 
= f_+(q^2) \left( p_B^\mu + p_\pi^\mu - \frac{m_B^2-m_\pi^2}{q^2}q^\mu\right)
+ f_0(q^2)\frac{m_B^2-m_\pi^2}{q^2}q^\mu.
\end{equation}
Nonperturbative lattice-QCD provides a first-principles method
for computing the $B \to \pi$ form factors with controlled uncertainties.

In these proceedings,
we report on a lattice-QCD calculation of the  $B \to \pi$ form factor with domain-wall light quarks 
using the $2 + 1$ flavor domain-wall fermion and Iwasaki gauge-field ensembles generated 
by the RBC and UKQCD Collaborations.
There have been  two  $2+1$ flavor lattice calculations of $f_+(q^2)$
done by the HPQCD~\cite{Dalgic:2006dt} and FNAL/MILC Collaborations~\cite{Bailey:2008wp}.
Both groups use the MILC gauge configurations.
Our calculation will provide an important independent 
check on existing calculations that use staggered light quarks.

\section{Methodology}
In practice, we  calculate the form factors $f_{\parallel}$ and $ f_{\perp}$,
which are more convenient for lattice simulations:
\begin{equation}
 \langle  \pi | \mathcal{V}^\mu | B \rangle = \sqrt{2m_B}
[v^\mu f_{\parallel}(E_\pi) + p^\mu_{\perp}f_{\perp}(E_\pi)],
\end{equation}
where $v^\mu = p_B^\mu/m_B$ and $p^\mu_{\perp}=p_\pi^\mu -(p_\pi\cdot v)v^\mu$.
In the $B$-meson rest frame, these form factors are
directly proportional to the hadronic matrix elements of the temporal and spatial vector current:
\begin{equation}
f_{\parallel} = \frac{\langle \pi | \mathcal{V}^0 |B\rangle }{\sqrt{2m_B}} , \ \ \ \ 
 f_{\perp} = \frac{\langle \pi | \mathcal{V}^i | B\rangle}{\sqrt{2m_B}} \frac{1}{p^i_\pi},
\end{equation}
and the desired form factors $f_+$ and $f_0$ can be obtained by the following relations:
\begin{eqnarray}  
 f_+(q^2) &=& \frac{1}{\sqrt{2m_B}}[f_{\parallel}(E_\pi)+(m_B-E_\pi)f_{\perp}(E_\pi)] \\
 f_0(q^2) &=& \frac{\sqrt{2m_B}}{m_B^2 - m_\pi^2}
              \left[
		(m_B-E_\pi) f_{\parallel}(E_\pi)+ (E_\pi^2 -m_\pi^2) f_{\perp}(E_\pi)
              \right].
\end{eqnarray}

\begin{table}[t]
  \centering
  \caption{Heavy-light current renormalization factors used in our analysis.
  The values of $\rho_{V_\mu}^{bl}$ are obtained in the chiral limit~\cite{Lehner:2012bt}.
  }
  \label{tableZ}
  \begin{tabular}{cccccc} \hline \hline
    & \multicolumn{2}{c}{$a \approx 0.11$~fm} &  \multicolumn{3}{c}{$a \approx 0.086$~fm} \\[-4pt] 
    & $am_l =0.005$ & $am_l=0.01$  & $am_l =0.004$ & $am_l=0.006$  & $am_l =0.008$ \\ \hline
    $Z_V^{ll}$   & 0.71732(14) & 0.71783(15) & 0.745053(54) & 0.745222(45) & 0.745328(48)\\ 
    $Z_V^{bb}$   & 10.037(34)  & 10.042(37)  & 5.270(13)    & 5.237(12)    &  5.267(15)  \\ \hline
    $\rho^{bl}_{V_0}$ & \multicolumn{2}{c}{1.02658} & \multicolumn{3}{c}{1.01661} \\
    $\rho^{bl}_{V_i}$ & \multicolumn{2}{c}{0.99723} & \multicolumn{3}{c}{0.99398} \\ \hline \hline
  \end{tabular}
\end{table}
We employ the  mostly nonperturbative method of Ref.~\cite{ElKhadra:2001rv} 
to match the lattice amplitude to the continuum matrix element:
\begin{equation}
\langle \pi | \mathcal{V}^\mu | B\rangle = Z_{V_\mu}^{bl} \langle \pi | V^\mu | B\rangle
 \ \ {\rm with} \ \ 
Z_{V_\mu}^{bl} = \rho_{V_\mu}^{bl} \sqrt{Z_{V}^{bb} Z_{V}^{ll} }.
\end{equation}
The flavor-conserving renormalization factors $Z_{V}^{bb}$ and  $ Z_{V}^{ll}$  are 
computed nonperturbatively 
on the lattice and the factor $\rho^{bl}_{V_\mu}$ is computed at one loop in mean-field improved
lattice perturbation theory~\cite{Lepage:1992xa}.
Most of the heavy-light current renormalization factor comes from $Z_V^{bb}$ and $Z_V^{ll}$, 
such that $\rho^{bl}_{V_\mu}$ is expected to be close to unity~\cite{Harada:2001fi}.
In this study, the factor $Z_V^{bb}$ is calculated  using 
the charge-normalization condition $Z_V^{bb}\langle B_s|V^{bb,0}|B_s\rangle =2m_B$
where $V^{bb,0}$ is the $b\to b$ lattice vector current.
We use the values of $Z_V^{ll}$ obtained by the RBC/UKQCD collaborations 
by exploiting the fact that $Z_A = Z_V$ for domain-wall fermions \cite{Aoki:2010dy}.
The renormalization factors used in this analysis are summarized in Table~\ref{tableZ}.

We improve the $b \to u$ vector current through $\mathcal{O}(\alpha_S a)$.  
At this order we need only compute one additional matrix element with a single-derivative operator.  
We calculate the improvement coefficient at 1-loop in mean-field improved lattice perturbation theory.  

\section{Computational setup}
\begin{table}[t]
  \centering
  \caption{ Lattice simulation parameters.
}
  \label{tab:ensembles}
 \begin{tabular}{ccccccc} \hline\hline 
  $a$ [fm] & $L^3\times T$ & $am_l$ & $am_s$ & $M_\pi$ [MeV] & \# configs. & \# time sources \\ \hline
  $\approx 0.11$ & $24^3\times 64$ &  0.005 &  0.040 &  329 &  1636 &  1 \\
  $\approx 0.11$ & $24^3\times 64$ &  0.010 &  0.040 &  422 &  1419 &  1 \\ 
  $\approx 0.086$& $32^3\times 64$ &  0.004 &  0.030 &  289 &  628 &  2 \\
  $\approx 0.086$ & $32^3\times 64$ & 0.006 & 0.030 & 345 & 889 & 2 \\ 
  $\approx 0.086$ & $32^3\times 64$ & 0.008 & 0.030 & 394 & 544 & 2 \\ \hline\hline

 \end{tabular}
\end{table}

We use the $2+1$ flavor domain-wall fermion and Iwasaki gauge-field ensembles
generated by the RBC and UKQCD Collaborations~\cite{Allton:2008pn,Aoki:2010dy}.
The simulation parameters are summarized in Table~\ref{tab:ensembles}.
We use the domain-wall action for the light valence-quark propagators, and unitary pion masses.
On the finer ensembles we compute two quark propagators on each configuration 
with their temporal source locations separated by $T/2$ to increase the statistics.
Periodic boundary conditions are imposed in the time direction.
Additional details of our computational setup are given in Ref.~\cite{Kawanai:2012id}.

For the bottom quark, we use the relativistic heavy quark (RHQ) action~\cite{Christ:2006us}
to control heavy-quark discretization errors introduced by the large bottom quark mass~\cite{ElKhadra:1996mp}.
For the parameters of the RHQ action, we need  calibration of  three parameters 
(the bare-quark mass $m_0 a$, the clover coefficient $c_P$, 
and the anisotropy parameter $\xi$)~\cite{Christ:2006us, Lin:2006ur}.
Here we employ values determined nonperturbatively  in Ref.~\cite{Aoki:2012xaa}.

\section{Analysis}

\subsection{Lattice form factors $f_{\parallel}^{\rm lat }$ and $ f_{\perp}^{\rm lat }$ }
The lattice form factors $f_{\parallel}^{\rm lat }$ and $ f_{\perp}^{\rm lat }$  
are obtained  from the following  ratios of correlation functions at large source-sink separation:
\begin{eqnarray}
R_{3,\mu}^{B\to \pi}(E_\pi, t,t_{\rm snk}) &=& \frac{C_{3,\mu}^{B\to \pi}(E_\pi, t,t_{\rm snk})}{\sqrt{C_2^{\pi}\
(E_\pi, t) C_2^{B}(t_{\rm snk}-t) }}
     \sqrt{\frac{2E_\pi}{e^{-E_\pi t}e^{-m_B (t_{\rm snk}-t)}}}.
\end{eqnarray}
%
%
 \begin{figure}[t] 
   \centering
   \includegraphics[width=.49\textwidth]{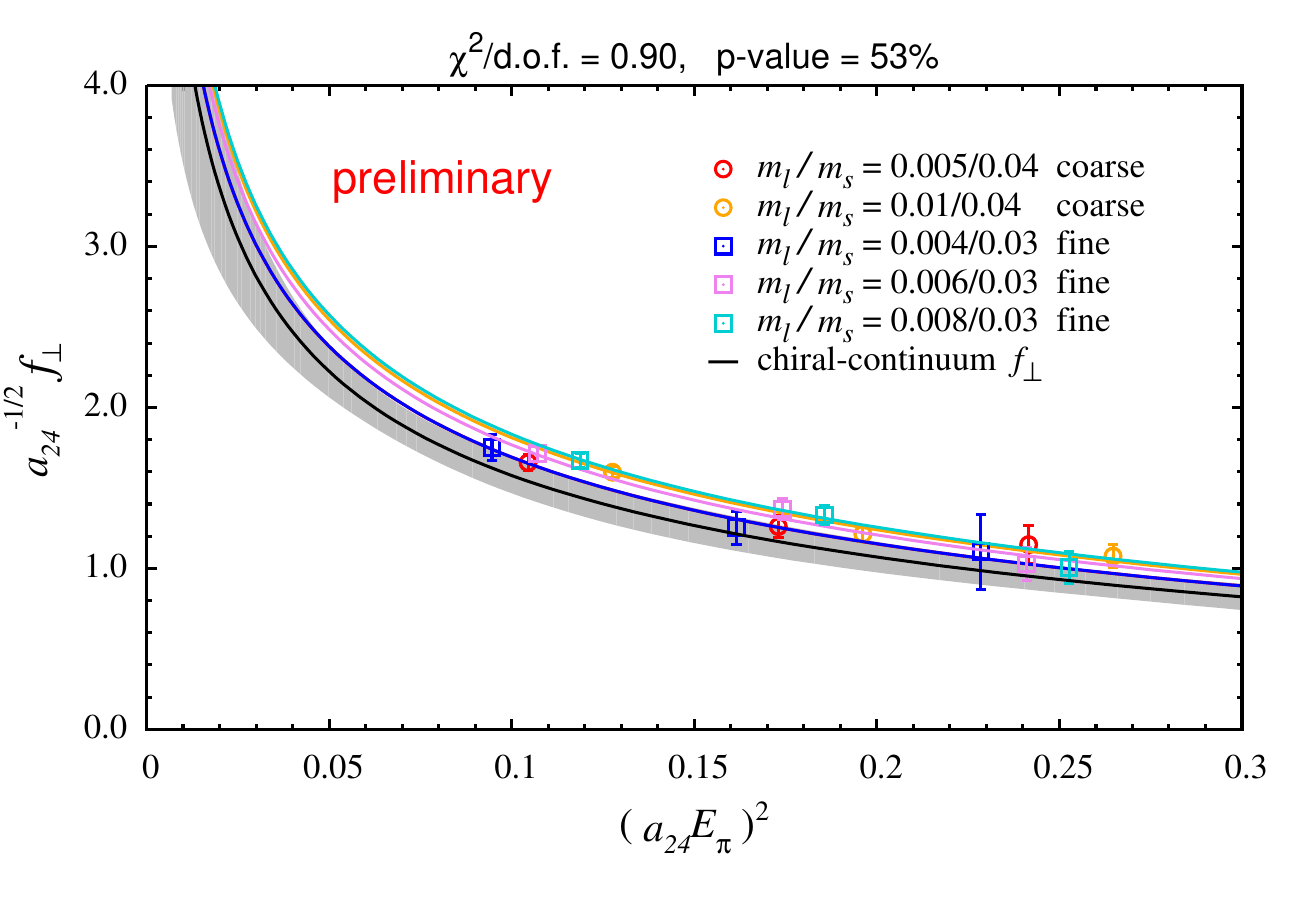}
   \includegraphics[width=.49\textwidth]{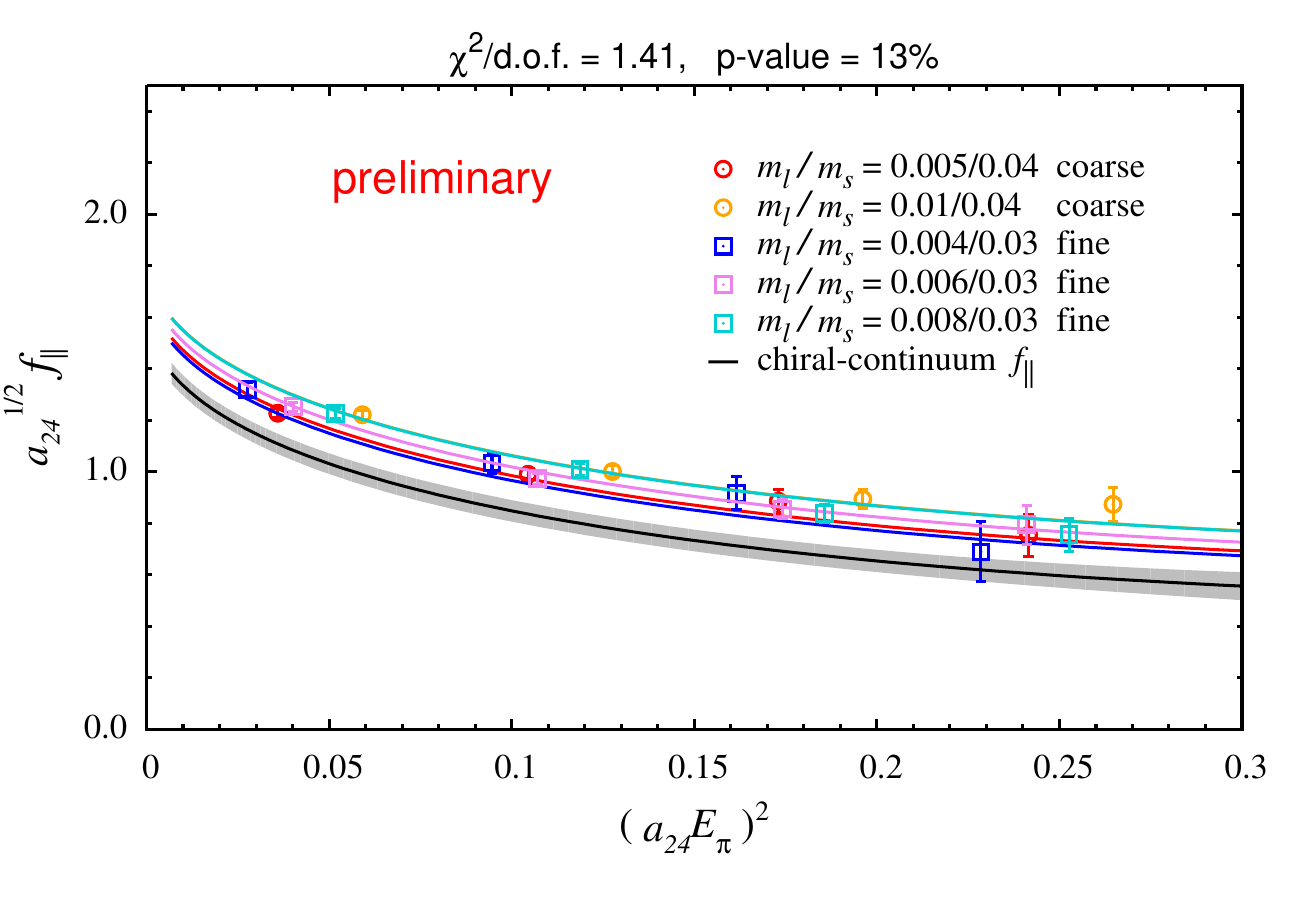}
   \caption{The $B \to \pi$ form factors $f_{\perp}$~(left) and $ f_{\para}$~(right),
    shown in units of the lattice spacing on the $24^3$ ensembles.
    The black curves with gray error bands show the
    chiral-continuum extrapolated $f_{\parallel}$ and $ f_{\perp}$ with statistical errors.
  }
   \label{fig:form}
\end{figure}
After multiplying the results for $R_{3,0}^{B\to \pi}$ and $R_{3,i}^{B\to \pi}$ by $Z_V^{bl}$:
\begin{eqnarray}
 f_{\perp}(E_\pi) &=& Z_{V_i}^{bl} \lim_{t,t_{\rm snk} \to \infty} \frac{1}{p_i} R_{3,i}^{B\to \pi} (E_\pi,t,t_{\rm snk}) \\
 f_{\para}(E_\pi) &=& Z_{V_0}^{bl} \lim_{t,t_{\rm snk} \to \infty}  R_{3,0}^{B\to \pi} (E_\pi,t,t_{\rm snk}),
\end{eqnarray}
we obtain the renormalized form factors $f_{\parallel}$ and $f_{\perp}$ as a function of pion energy on each ensemble, 
shown in Fig.~\ref{fig:form}.
We use the form factor data through momentum $\vec{p} = 2\pi (1,1,1)/L$ in this analysis.

\subsection{Chiral-continuum extrapolation}
In order to extrapolate simultaneously the form factors to the physical light-quark mass and the continuum,
we employ NLO $SU(2)$ hard-pion chiral perturbation theory~\cite{Bijnens:2010ws}
suitably modified to incorporate leading discretization errors from the domain-wall and Iwasaki actions.
Thus the fit functions depend on the pion mass $m_{ll}$, pion-energy $E_\pi$ and squared lattice spacing $a^2$:
\begin{eqnarray}
 f_\perp(m_{ll}, E_\pi, a^2) &=&  
  \frac{c_\perp^{(1)}}{E_\pi+\Delta}
  \left(1+ \delta f_\perp
  + c_\perp^{(2)} m_{ll}^2  
  + c_\perp^{(4)} E_\pi  + c_\perp^{(5)} E_\pi^2
  + c_\perp^{(6)} a^2  \right) \\
 f_\para(m_{ll}, E_\pi, a^2) &=&  
  c_\para^{(1)}\left(1+ \delta f_\para
  + c_\para^{(2)} m_{ll}^2 
  + c_\para^{(4)} E_\pi  + c_\para^{(5)} E_\pi^2
  + c_\para^{(6)} a^2  \right),
\end{eqnarray}
where the quantity $\Delta$ is the mass difference $m_{B^*} - m_B$, fixed to the experimental value~\cite{Beringer:1900zz}, and 
ensures the proper location of the pole at the $B^*$ mass in the physical form factor $f_+$.
The function $\delta f$ contains logarithmic functions of the pion mass:
\begin{eqnarray}
  (4\pi f_\pi)^2 \delta f_{\para/\perp}
 &=& -\frac{9g^2}{4} I_1(m_{ll})
     - I_1(m_{ll}) + \frac{1}{4}I_1(m_{ll})
    +\frac{1}{4} (m_{ll}^2 - m_{ll}^2)\frac{\partial I_1(m_{ll})}{\partial m_{ll}^2},
\end{eqnarray}
where $I_1(m  _{ll}) = m_{ll}^2 \log(m_{ll}^2/\Lambda^2)$ and $f_\pi = 130.4$~MeV~\cite{Beringer:1900zz}.
For the parameter $g$, we use $g_{B^*B \pi} = 0.569$ calculated 
as a part of this RHQ project~\cite{Ben:lattice2013}.
Fig.~\ref{fig:form} shows the resulting chiral-continuum extrapolation of $f_\perp$ and $f_\para$.

\subsection{Extrapolation in $q^2$ to zero recoil}
\begin{figure}[t] 
  \centering
  \includegraphics[width=.49\textwidth]{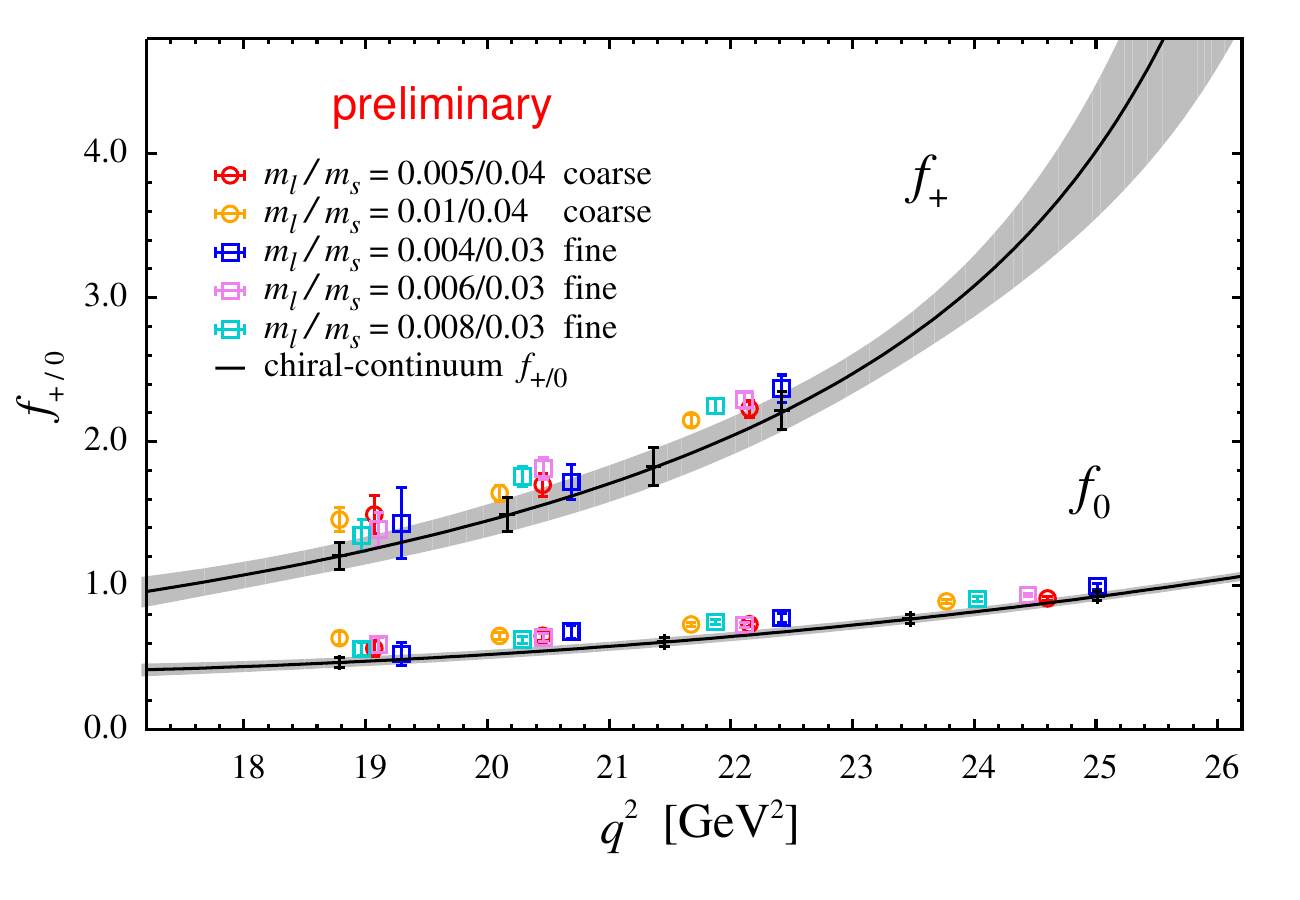}
  \caption{The $B \to \pi$ form factors $f_+$ and $f_0$
    The black curves with gray error bands show the chiral-continuum extrapolated form factors with statistical errors
  with the four evenly-spaced synthetic data points used in the $q^2$ extrapolation overlaid.  Errors shown are statistical only.}
   \label{fig:form2}
\end{figure}
We must extrapolate the lattice data to lower $q^2$ (larger $E_\pi^2$ ) 
to reach the kinematic region where experimental measurements are most precise.
Using chiral-continuum extrapolated lattice data
in the range of simulated pion energies, we first generate four synthetic data points of 
the form factors $f_+$ and $f_0$ used in the $q^2$ extrapolation to the full kinematic range,
as shown in Fig.~\ref{fig:form2}.

In this study, we employ the model-independent $z$-expansion fit 
to extrapolate to low momentum transfer~\cite{Bourrely:1980gp,Boyd:1994tt,Arnesen:2005ez,Bourrely:2008za}.
As a first step, we consider mapping the variable $q^2$ on to a new variable $z$ defined as 
\begin{equation}
 z=\frac{\sqrt{t_+ - q^2} - \sqrt{t_+ -t_0}}{\sqrt{t_+ - q^2} + \sqrt{t_+ - t_0}}
\end{equation}
where $t_{\pm} = (m_B \pm m_\pi)^2$.
This transformation maps the semileptonic region $0 < q^2 < t_-$  
onto small values of $z$ between $−0.34<z<0.22$ when we choose $t_0 =0.65 t_+$.
The $B \to \pi$ form factors are analytic in the semileptonic region except at the location of the $B^*$ pole,
so the form factors $f_+$ and $f_0$ can be expressed as convergent power series:
\begin{equation}
  f(q^2) = \frac{1}{P(q^2)\phi(q^2,t_0)} \sum_{k=0}^{\infty} a^{(k)} (t_0) z(q^2, t_0)^k,
\end{equation}
where the function $P(q^2)$ is the Blaschke factor that contains subthreshold poles,
 and the outer function $\phi(q^2,t_0)$ is an arbitrary analytic function.
Unitarity constrains the sum of the squares of the coefficients 
because the decay process of $B \to \pi l \nu$ semileptonic decay is 
related to the scattering process $l\nu \to B \pi$ by crossing symmetry.
When the outer function is chosen as in Ref.~\cite{Boyd:1994tt}, 
the sum of the squares of the coefficients is bounded by unity: $\sum_{k=0}^{N} (a^{(k)})^2 \leq 1$ for any $N$.
Therefore only a small number of terms is needed to accurately describe the shape of the form factors over the full kinematic range.

\begin{figure}[t] 
  \centering
  \includegraphics[width=.49\textwidth]{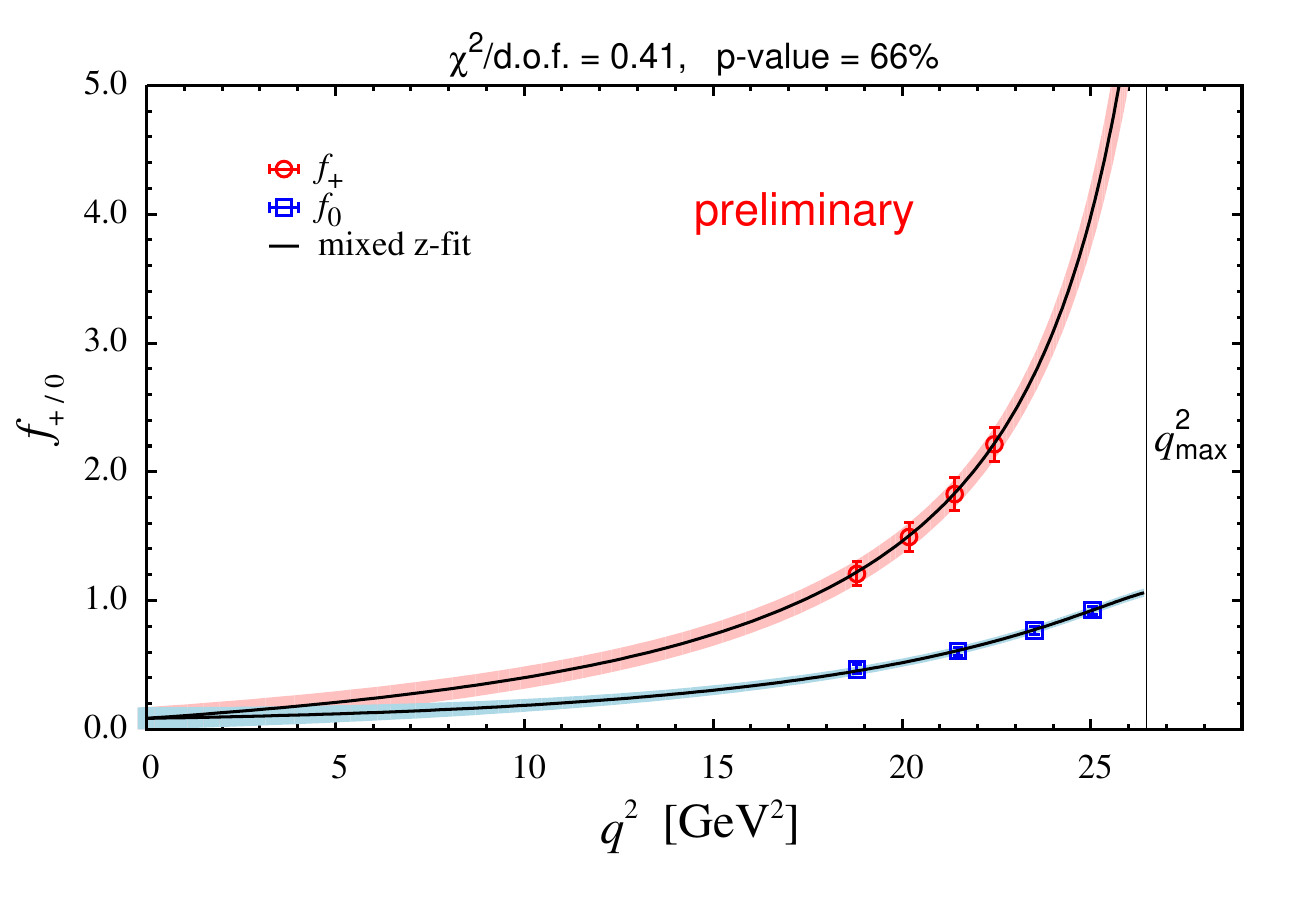}
  \includegraphics[width=.49\textwidth]{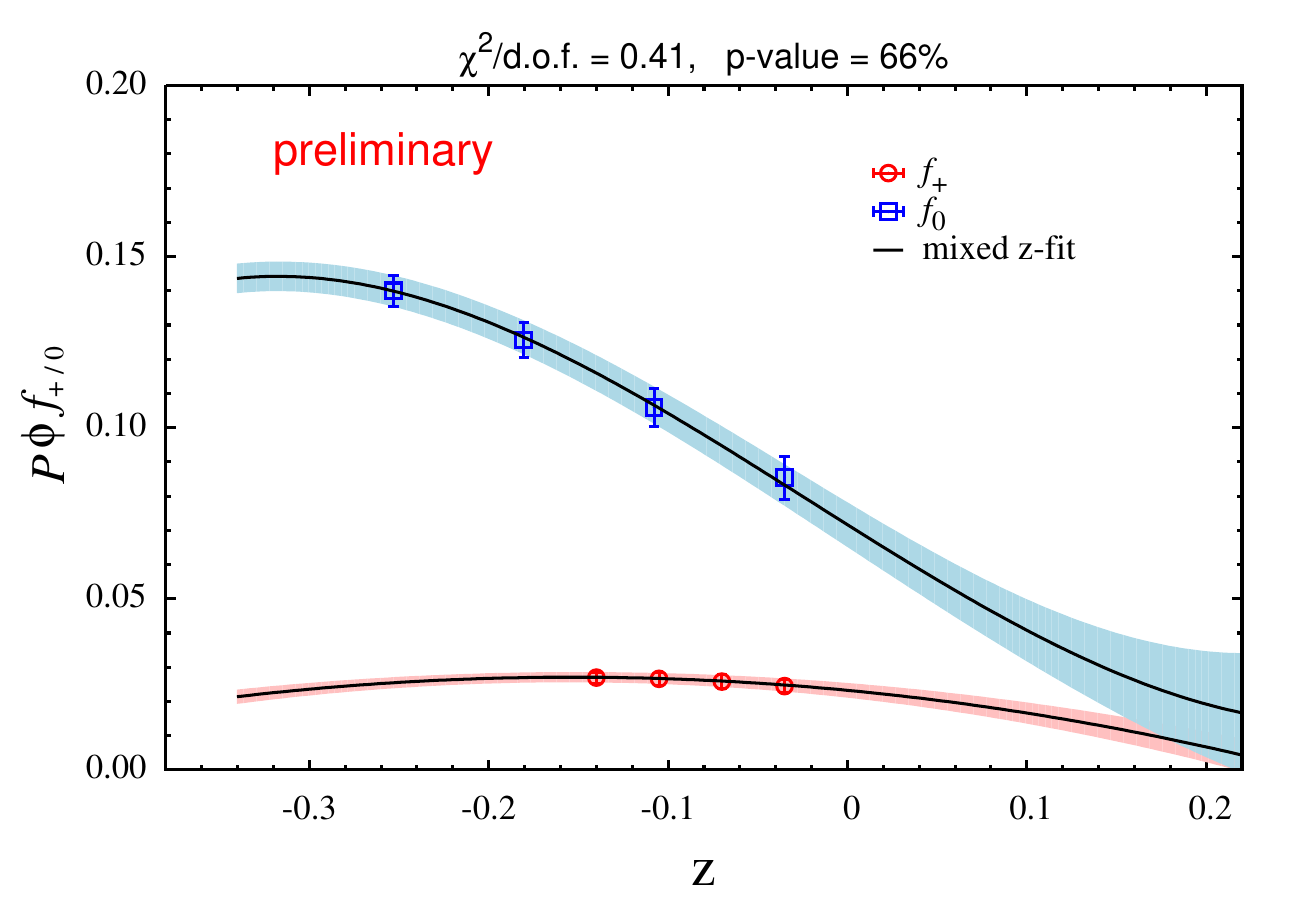}
  \caption{Extrapolation in  $q^2$ of the factors $f_+$ and $f_0$ 
    using the model-independent BGL $z$-parametrization, and imposing the kinematic constraint $f_+(q^2=0) = f_0(q^2=0)$.
  The left plot shows $f_{+/0}$ vs. $q^2$, while the right plot shows $P\phi f_{+/0}$ vs. $z$.
   The black curves with  error bands show the parametrization of 
  the form factors over the full kinematic range.  Errors are statistical only.}
   \label{fig:z_fit}
\end{figure}
Fig.~\ref{fig:z_fit} shows fits of the form factors $f_+$ and $f_0$
using the $z$-parametrization in Boyd, Grinstein, and Lebed (BGL)~\cite{Boyd:1994tt}.
In the fits, we includes terms up to $z^2$ for $f_+$ and $z^3$ for $f_0$, and 
the kinematic constraint $f_+(q^2=0) = f_0(q^2=0)$ is imposed.
The resulting slope and curvature  for the $B \to \pi l \nu$ vector form factor $f_+$ are 
\begin{eqnarray}
 a_+^{(1)}/ a_+^{(0)} &=& -2.15 \pm 0.63 \\
 a_+^{(2)}/ a_+^{(0)} &=& -7.02 \pm 1.20,
\end{eqnarray}
where the errors are from statistics only.
Our preliminary  values of the slope and curvature are consistent with the
 independent lattice determination from the Fermilab Lattice and MILC Collaborations~\cite{Bailey:2008wp}.
%

\section{Outlook}
We are currently estimating the systematic uncertainties in the form factors $f_+$ and $f_0$. 
 We expect the dominant source of error to be from the chiral-continuum extrapolation, 
and that our total error will be competitive with that of Ref.~\cite{Bailey:2008wp}.  

Once we have a complete error budget, 
we will extrapolate our results from the simulated pion energies 
to the full $q^2$ range using the $z$-parametrization.  
Currently we are using the $z$-parametrization of Boyd, Grinstein, and Lebed, 
but we will also consider the alternative parametrization of 
Bourrely, Caprini, and Lellouch~\cite{Bourrely:2008za}.  
We will perform the $z$-fit to the lattice data alone to provide 
a model-independent parametrization of our result valid 
over the full kinematic range.  
We will also perform a simultaneous $z$-fit of our data and 
experimental measurements of the $B\to \pi$ differential branching fraction to obtain $|V_{ub}|$.  

Our results will provide an important independent check on existing calculations, all of which use staggered light quarks.

\section{Acknowledgments}
The authors wish to thank our collaborators in the RBC and UKQCD Collaborations
for helpful discussions.
Computations for this work were mainly performed on resources provided
by the USQCD Collaboration, funded by the Office of Science of the U.S. Department of
Energy, as well as computers at BNL and Columbia University.
T. Kawanai was partially supported by JSPS Strategic Young Researcher Overseas Visits Program
  for Accelerating Brain Circulation (No.~R2411).
O.W. acknowledges support at Boston University by the U.S. DOE grant DE-SC0008814.
BNL is operated by Brookhaven Science Associates, LLC under Contract 
No. DE-AC02-98CH10886 with the U.S. Department of Energy.  
Fermilab is operated by Fermi Research Alliance, LLC, under Contract 
No. DE-AC02-07CH11359 with the U.S. Department of Energy.

{\small
\bibliographystyle{apsrev4-1}
\bibliography{lattice2013}
}

\end{document}